\documentclass{article}[10pt,letterpaper]
\usepackage{fancyhdr}
\usepackage{supertabular}
\usepackage{color}
\usepackage{amsmath}
\usepackage{pbox}
\usepackage{amssymb}
\usepackage{amsbsy}
\usepackage{scalefnt}
\usepackage{ifpdf}
\usepackage{graphicx}
\usepackage{multicol}
\usepackage[colorlinks=true, linkcolor=blue, linktocpage=true,urlcolor=blue]{hyperref}
\usepackage{longtable}
\usepackage{needspace}
\usepackage{eso-pic}
\usepackage{mathptmx}
\usepackage{multirow}
\usepackage{lscape}
\usepackage{textcomp}
\usepackage{changepage}
\usepackage{xfrac}
\usepackage{txfonts}
\usepackage{pdfcomment}
\usepackage[absolute]{textpos}
\ifpdf
\fi
\makeatletter
\renewcommand{\l@section}{\@dottedtocline{2}{1em}{0em}}
\renewcommand{\l@subsection}{\@dottedtocline{3}{2.5cm}{0em}}
\renewcommand{\hrulefill}{\leavevmode \leaders \hrule \@height 1pt \hfill \kern\z@}
\makeatother
\renewcommand{\underline}[1]{\begin{tabular}{@{\extracolsep{\fill}}c@{\extracolsep{\fill}}}#1\\[-0.2cm]\hrulefill\end{tabular}}

\newcounter{chappage}
\AddToShipoutPicture{\stepcounter{chappage}}
\fancypagestyle{plain}{\lhead{}\rhead{}}
\fancypagestyle{single}{\lhead{\leftmark \ \\}\rhead{\leftmark \ \\}}
\fancypagestyle{bob}{\lhead{\leftmark -\thechappage \ \\}\rhead{\leftmark -\thechappage \ \\}}
\begin{document}
\fontsize{9}{10}\selectfont
\fontdimen2\font=1.3\fontdimen2\font
\thispagestyle{empty}
\setcounter{page}{1}
\begin{center}

\vspace{0.5cm}
{ \huge Energy Levels of \ensuremath{^{\textnormal{20}}}Al*}\\
\vspace{1.0cm}
{ \normalsize K. Setoodehnia\ensuremath{^{\textnormal{1}}} and J. H. Kelley\ensuremath{^{\textnormal{1,2}}}}\\
\vspace{0.2in}
{ \small \it \ensuremath{^{\textnormal{1}}}Triangle Universities Nuclear Laboratory, Duke University,\\
  Durham North Carolina 27708, USA.\\
  \ensuremath{^{\textnormal{2}}}Department of Physics, North Carolina State University\\
  Raleigh, North Carolina 27607, USA}\\
\vspace{0.2in}
\end{center}

\setlength{\parindent}{-0.5cm}
\addtolength{\leftskip}{2cm}
\addtolength{\rightskip}{2cm}
{\bf Abstract: }
Experimental nuclear structure data are evaluated for \ensuremath{^{\textnormal{20}}}Al. The 3p-unbound, previously unobserved \ensuremath{^{\textnormal{20}}}Al nucleus was recently discovered by (\href{https://www.nndc.bnl.gov/nsr/nsrlink.jsp?2025Xu03,B}{2025Xu03}) through the analysis of the (\href{https://www.nndc.bnl.gov/nsr/nsrlink.jsp?2007Mu15,B}{2007Mu15}: \ensuremath{^{\textnormal{9}}}Be(\ensuremath{^{\textnormal{20}}}Mg,{\allowbreak}\ensuremath{^{\textnormal{19}}}Mg)) data. (\href{https://www.nndc.bnl.gov/nsr/nsrlink.jsp?2025Xu03,B}{2025Xu03}) analyzed the by-product \ensuremath{^{\textnormal{9}}}Be(\ensuremath{^{\textnormal{20}}}Mg,{\allowbreak}\ensuremath{^{\textnormal{20}}}Al) charge exchange reaction channel followed by an in-flight decay of \ensuremath{^{\textnormal{20}}}Al. \ensuremath{^{\textnormal{20}}}Al\ensuremath{_{\textnormal{g.s.}}} decays by one-proton emission to \ensuremath{^{\textnormal{19}}}Mg\ensuremath{_{\textnormal{g.s.}}}, which is also particle unbound and democratically decays to \ensuremath{^{\textnormal{17}}}Ne\ensuremath{_{\textnormal{g.s.}}}+2p.\\

{\bf Cutoff Date: }
Literature available up to November 21, 2025 has been considered. The primary bibliographic source is the NSR database (\href{https://www.nndc.bnl.gov/nsr/nsrlink.jsp?2011Pr03,B}{2011Pr03}) available at Brookhaven National Laboratory web page: www.nndc.bnl.gov/nsr/.\\

{\bf General Policies and Organization of Material: }
See the April 2026 issue of the {\it Nuclear Data Sheets} or \\https://www.nndc.bnl.gov/nds/docs/NDSPolicies.pdf. \\

{\bf Acknowledgements: }
The authors expresses her gratitude to personnel at the National Nuclear Data Center (NNDC) at Brookhaven National Laboratory for facilitating this work.\\

\vfill

* This work is supported by the Office of Nuclear Physics, Office of Science, U.S. Department of Energy under contracts: DE-FG02-97ER41042 {\textminus} North Carolina State University and DE-FG02-97ER41033 {\textminus} Duke University\\

\setlength{\parindent}{+0.5cm}
\addtolength{\leftskip}{-2cm}
\addtolength{\rightskip}{-2cm}
\newpage
\pagestyle{plain}
\clearpage
\pagestyle{bob}
\begin{center}
\section[\ensuremath{^{20}_{13}}Al\ensuremath{_{7}^{~}}]{ }
\vspace{-30pt}
\setcounter{chappage}{1}
\subsection[\hspace{-0.2cm}Adopted Levels]{ }
\vspace{-20pt}
\vspace{0.3cm}
\hypertarget{AL0}{{\bf \small \underline{Adopted \hyperlink{20AL_LEVEL}{Levels}}}}\\
\vspace{4pt}
\vspace{8pt}
\parbox[b][0.3cm]{17.7cm}{\addtolength{\parindent}{-0.2in}S(p)=1170 {\it 13}\hspace{0.2in}\href{https://www.nndc.bnl.gov/nsr/nsrlink.jsp?2025Xu03,B}{2025Xu03},\href{https://www.nndc.bnl.gov/nsr/nsrlink.jsp?2021Wa16,B}{2021Wa16}}\\
\parbox[b][0.3cm]{17.7cm}{\addtolength{\parindent}{-0.2in}S(p): From S\ensuremath{_{\textnormal{p}}}=1.17 MeV \textit{+10{\textminus}8} (\href{https://www.nndc.bnl.gov/nsr/nsrlink.jsp?2025Xu03,B}{2025Xu03}), which is deduced by those authors from S\ensuremath{_{\textnormal{3p}}}=1.17 MeV \textit{+10{\textminus}8} measured by}\\
\parbox[b][0.3cm]{17.7cm}{(\href{https://www.nndc.bnl.gov/nsr/nsrlink.jsp?2025Xu03,B}{2025Xu03}).}\\
\parbox[b][0.3cm]{17.7cm}{\addtolength{\parindent}{-0.2in}(\href{https://www.nndc.bnl.gov/nsr/nsrlink.jsp?2025Xu03,B}{2025Xu03}) deduced \ensuremath{\Delta}M=40.30 MeV \textit{12} (mass excess) from S\ensuremath{_{\textnormal{3p}}}(\ensuremath{^{\textnormal{20}}}Al\ensuremath{_{\textnormal{g.s.}}})=1.93 MeV \textit{+12{\textminus}10} and the mass excesses of proton}\\
\parbox[b][0.3cm]{17.7cm}{and \ensuremath{^{\textnormal{17}}}Ne from (\href{https://www.nndc.bnl.gov/nsr/nsrlink.jsp?2021Wa16,B}{2021Wa16}).}\\

\parbox[b][0.3cm]{17.7cm}{\addtolength{\parindent}{-0.2in}The \ensuremath{^{\textnormal{20}}}Al nucleus is not mentioned in the most recent AME-2020 analysis (\href{https://www.nndc.bnl.gov/nsr/nsrlink.jsp?2021Wa16,B}{2021Wa16}).}\\
\vspace{0.385cm}
\parbox[b][0.3cm]{17.7cm}{\addtolength{\parindent}{-0.2in}\textit{Theoretical Analyses}:}\\
\parbox[b][0.3cm]{17.7cm}{\addtolength{\parindent}{-0.2in}\href{https://www.nndc.bnl.gov/nsr/nsrlink.jsp?1987Po01,B}{1987Po01}: Calculated ground state spin, \textit{sd} configuration percentage, and correlation energy using Shell model.}\\
\parbox[b][0.3cm]{17.7cm}{\addtolength{\parindent}{-0.2in}\href{https://www.nndc.bnl.gov/nsr/nsrlink.jsp?1997Ki22,B}{1997Ki22}: Calculated \textit{rms} radii and their related features, mass quadrupole moment, and density contour using Hartree-Fock model}\\
\parbox[b][0.3cm]{17.7cm}{with the SGII force; comparison with mirror nucleus, \ensuremath{^{\textnormal{20}}}N.}\\
\parbox[b][0.3cm]{17.7cm}{\addtolength{\parindent}{-0.2in}\href{https://www.nndc.bnl.gov/nsr/nsrlink.jsp?1999Kn04,B}{1999Kn04}: Calculated radii using a phenomenological, semi-microscopic approach.}\\
\parbox[b][0.3cm]{17.7cm}{\addtolength{\parindent}{-0.2in}\href{https://www.nndc.bnl.gov/nsr/nsrlink.jsp?2013Ti01,B}{2013Ti01}: Calculated binding energy, mass excess, S\ensuremath{_{\textnormal{p}}}, S\ensuremath{_{\textnormal{2p}}} using improved Kelson-Garvey (ImKG) mass relations. Comparison}\\
\parbox[b][0.3cm]{17.7cm}{with experimental data for mirror analogs. Prediction of mass. Discussed diproton emission.}\\
\parbox[b][0.3cm]{17.7cm}{\addtolength{\parindent}{-0.2in}\href{https://www.nndc.bnl.gov/nsr/nsrlink.jsp?2018Fo04,B}{2018Fo04}: Analyzed and fitted mirror energy differences (MED) between 2\textit{s}\ensuremath{_{\textnormal{1/2}}} and 1\textit{d}\ensuremath{_{\textnormal{5/2}}} single-particle states in \ensuremath{^{\textnormal{20}}}Al and \ensuremath{^{\textnormal{20}}}N}\\
\parbox[b][0.3cm]{17.7cm}{mirrors. Comparison with improved Kelson-Garvey (ImKG) model.}\\
\parbox[b][0.3cm]{17.7cm}{\addtolength{\parindent}{-0.2in}\href{https://www.nndc.bnl.gov/nsr/nsrlink.jsp?2022Zo01,B}{2022Zo01}: Calculated S\ensuremath{_{\textnormal{p}}}, S\ensuremath{_{\textnormal{2p}}}, mass excess, in terms of mass relations for mirror nuclei, based on Weizsacker mass formula.}\\
\vspace{0.385cm}
\vspace{12pt}
\hypertarget{20AL_LEVEL}{\underline{$^{20}$Al Levels}}\\
\begin{longtable}[c]{ll}
\multicolumn{2}{c}{\underline{Cross Reference (XREF) Flags}}\\
 \\
\hyperlink{AL1}{\texttt{A }}& \ensuremath{^{\textnormal{9}}}Be(\ensuremath{^{\textnormal{20}}}Mg,\ensuremath{^{\textnormal{20}}}Al)\\
\end{longtable}
\vspace{-0.5cm}
\begin{longtable}{ccccccccc@{\extracolsep{\fill}}c}
\multicolumn{2}{c}{E(level)$^{{\hyperlink{AL0LEVEL0}{a}}}$}&J$^{\pi}$$^{{\hyperlink{AL0LEVEL1}{b}}}$&\multicolumn{2}{c}{\ensuremath{\Gamma}$^{{\hyperlink{AL0LEVEL0}{a}}}$}&\multicolumn{2}{c}{E\ensuremath{_{\textnormal{c.m.}}}(\ensuremath{^{\textnormal{17}}}Ne+3p) (MeV)$^{{\hyperlink{AL0LEVEL0}{a}}}$}&XREF&Comments&\\[-.2cm]
\multicolumn{2}{c}{\hrulefill}&\hrulefill&\multicolumn{2}{c}{\hrulefill}&\multicolumn{2}{c}{\hrulefill}&\hrulefill&\hrulefill&
\endfirsthead
\multicolumn{1}{r@{}}{0}&\multicolumn{1}{@{}l}{}&\multicolumn{1}{l}{(1\ensuremath{^{-}})}&\multicolumn{1}{r@{}}{$<$400}&\multicolumn{1}{@{ }l}{keV}&\multicolumn{1}{r@{}}{1}&\multicolumn{1}{@{.}l}{93 {\it 16}}&\multicolumn{1}{l}{\texttt{\hyperlink{AL1}{A}} }&\parbox[t][0.3cm]{7.46865cm}{\raggedright \%p=100 (\href{https://www.nndc.bnl.gov/nsr/nsrlink.jsp?2025Xu03,B}{2025Xu03})\vspace{0.1cm}}&\\
&&&&&&&&\parbox[t][0.3cm]{7.46865cm}{\raggedright \ensuremath{\Gamma}: Mainly due to the experimental resolution (see\vspace{0.1cm}}&\\
&&&&&&&&\parbox[t][0.3cm]{7.46865cm}{\raggedright {\ }{\ }{\ }Appendix C in \href{https://www.nndc.bnl.gov/nsr/nsrlink.jsp?2025Xu03,B}{2025Xu03}).\vspace{0.1cm}}&\\
&&&&&&&&\parbox[t][0.3cm]{7.46865cm}{\raggedright E\ensuremath{_{\textnormal{c.m.}}}(\ensuremath{^{\textnormal{17}}}Ne+3p) (MeV): From S\ensuremath{_{\textnormal{3p}}}(\ensuremath{^{\textnormal{20}}}Al\ensuremath{_{\textnormal{g.s.}}})=1.93\vspace{0.1cm}}&\\
&&&&&&&&\parbox[t][0.3cm]{7.46865cm}{\raggedright {\ }{\ }{\ }MeV \textit{+12{\textminus}10} (\href{https://www.nndc.bnl.gov/nsr/nsrlink.jsp?2025Xu03,B}{2025Xu03}).\vspace{0.1cm}}&\\
&&&&&&&&\parbox[t][0.3cm]{7.46865cm}{\raggedright S\ensuremath{_{\textnormal{p}}}(\ensuremath{^{\textnormal{20}}}Al\ensuremath{_{\textnormal{g.s.}}})=1.17 MeV \textit{+10{\textminus}8} (\href{https://www.nndc.bnl.gov/nsr/nsrlink.jsp?2025Xu03,B}{2025Xu03}: See\vspace{0.1cm}}&\\
&&&&&&&&\parbox[t][0.3cm]{7.46865cm}{\raggedright {\ }{\ }{\ }Appendix A). This value corresponds to a\vspace{0.1cm}}&\\
&&&&&&&&\parbox[t][0.3cm]{7.46865cm}{\raggedright {\ }{\ }{\ }probability of 0.821 from a Kolmogorov test.\vspace{0.1cm}}&\\
&&&&&&&&\parbox[t][0.3cm]{7.46865cm}{\raggedright Decay mode: p+\ensuremath{^{\textnormal{19}}}Mg\ensuremath{_{\textnormal{g.s.}}}, where \ensuremath{^{\textnormal{19}}}Mg\ensuremath{_{\textnormal{g.s.}}}\vspace{0.1cm}}&\\
&&&&&&&&\parbox[t][0.3cm]{7.46865cm}{\raggedright {\ }{\ }{\ }democratically decays to \ensuremath{^{\textnormal{17}}}Ne+2p (\href{https://www.nndc.bnl.gov/nsr/nsrlink.jsp?2025Xu03,B}{2025Xu03}).\vspace{0.1cm}}&\\
&&&&&&&&\parbox[t][0.3cm]{7.46865cm}{\raggedright Configuration: \ensuremath{^{\textnormal{19}}}Mg+p, where the proton is\vspace{0.1cm}}&\\
&&&&&&&&\parbox[t][0.3cm]{7.46865cm}{\raggedright {\ }{\ }{\ }predominantly in the \textit{s}\ensuremath{_{\textnormal{1/2}}} orbital (\href{https://www.nndc.bnl.gov/nsr/nsrlink.jsp?2025Xu03,B}{2025Xu03}).\vspace{0.1cm}}&\\
\multicolumn{1}{r@{}}{1.67\ensuremath{\times10^{3}}}&\multicolumn{1}{@{ }l}{{\it 27}}&\multicolumn{1}{l}{(2\ensuremath{^{-}})}&&&\multicolumn{1}{r@{}}{3}&\multicolumn{1}{@{.}l}{60 {\it 22}}&\multicolumn{1}{l}{\texttt{\hyperlink{AL1}{A}} }&\parbox[t][0.3cm]{7.46865cm}{\raggedright \%p=100 (\href{https://www.nndc.bnl.gov/nsr/nsrlink.jsp?2025Xu03,B}{2025Xu03})\vspace{0.1cm}}&\\
&&&&&&&&\parbox[t][0.3cm]{7.46865cm}{\raggedright Decays to \ensuremath{^{\textnormal{19}}}Mg*(1.38 MeV, (3/2\ensuremath{^{-}}))+p, where the\vspace{0.1cm}}&\\
&&&&&&&&\parbox[t][0.3cm]{7.46865cm}{\raggedright {\ }{\ }{\ }proton energy is deduced to be E\ensuremath{_{\textnormal{c.m.}}}=1.50 MeV \textit{10}.\vspace{0.1cm}}&\\
&&&&&&&&\parbox[t][0.3cm]{7.46865cm}{\raggedright {\ }{\ }{\ }The \ensuremath{^{\textnormal{19}}}Mg*(1.38 MeV) level sequentially decays via\vspace{0.1cm}}&\\
&&&&&&&&\parbox[t][0.3cm]{7.46865cm}{\raggedright {\ }{\ }{\ }\ensuremath{^{\textnormal{18}}}Na(g.s., (1\ensuremath{^{-}})) to \ensuremath{^{\textnormal{17}}}Ne\ensuremath{_{\textnormal{g.s.}}}+2p.\vspace{0.1cm}}&\\
\end{longtable}
\parbox[b][0.3cm]{17.7cm}{\makebox[1ex]{\ensuremath{^{\hypertarget{AL0LEVEL0}{a}}}} From (\href{https://www.nndc.bnl.gov/nsr/nsrlink.jsp?2025Xu03,B}{2025Xu03}).}\\
\parbox[b][0.3cm]{17.7cm}{\makebox[1ex]{\ensuremath{^{\hypertarget{AL0LEVEL1}{b}}}} From Gamow shell model and Gamow coupled-channel model calculations in (\href{https://www.nndc.bnl.gov/nsr/nsrlink.jsp?2025Xu03,B}{2025Xu03}), where the ground and first excited}\\
\parbox[b][0.3cm]{17.7cm}{{\ }{\ }states were firmly and tentatively assigned, respectively. However, the evaluator made the ground state assignment tentative.}\\
\vspace{0.5cm}
\clearpage
\subsection[\hspace{-0.2cm}\ensuremath{^{\textnormal{9}}}Be(\ensuremath{^{\textnormal{20}}}Mg,\ensuremath{^{\textnormal{20}}}Al)]{ }
\vspace{-27pt}
\vspace{0.3cm}
\hypertarget{AL1}{{\bf \small \underline{\ensuremath{^{\textnormal{9}}}Be(\ensuremath{^{\textnormal{20}}}Mg,\ensuremath{^{\textnormal{20}}}Al)\hspace{0.2in}\href{https://www.nndc.bnl.gov/nsr/nsrlink.jsp?2025Xu03,B}{2025Xu03}}}}\\
\vspace{4pt}
\vspace{8pt}
\parbox[b][0.3cm]{17.7cm}{\addtolength{\parindent}{-0.2in}Charge exchange reaction.}\\
\parbox[b][0.3cm]{17.7cm}{\addtolength{\parindent}{-0.2in}J\ensuremath{^{\ensuremath{\pi}}}(\ensuremath{^{\textnormal{9}}}Be\ensuremath{_{\textnormal{g.s.}}})=3/2\ensuremath{^{-}} and J\ensuremath{^{\ensuremath{\pi}}}(\ensuremath{^{\textnormal{20}}}Mg\ensuremath{_{\textnormal{g.s.}}})=0\ensuremath{^{\textnormal{+}}}.}\\
\parbox[b][0.3cm]{17.7cm}{\addtolength{\parindent}{-0.2in}\href{https://www.nndc.bnl.gov/nsr/nsrlink.jsp?2025Xu03,B}{2025Xu03}: This study is credited for the discovery of \ensuremath{^{\textnormal{20}}}Al isotope. The authors reanalyzed the data of (\href{https://www.nndc.bnl.gov/nsr/nsrlink.jsp?2007Mu15,B}{2007Mu15}) by}\\
\parbox[b][0.3cm]{17.7cm}{investigating the \ensuremath{^{\textnormal{9}}}Be(\ensuremath{^{\textnormal{20}}}Mg,\ensuremath{^{\textnormal{20}}}Al) charge exchange reaction channel using the 4-fold \ensuremath{^{\textnormal{17}}}Ne+3p coincidence events from the in{\textminus}flight}\\
\parbox[b][0.3cm]{17.7cm}{decay of \ensuremath{^{\textnormal{20}}}Al.}\\
\parbox[b][0.3cm]{17.7cm}{\addtolength{\parindent}{-0.2in}The original experiment of (\href{https://www.nndc.bnl.gov/nsr/nsrlink.jsp?2007Mu15,B}{2007Mu15}) was carried out at the GSI/FRS facility, where a 450-MeV/nucleon \ensuremath{^{\textnormal{20}}}Mg beam was}\\
\parbox[b][0.3cm]{17.7cm}{produced from fragmentation of a 591-MeV/nucleon \ensuremath{^{\textnormal{24}}}Mg beam on a thick \ensuremath{^{\textnormal{9}}}Be target. The \ensuremath{^{\textnormal{20}}}Mg beam was delivered to a}\\
\parbox[b][0.3cm]{17.7cm}{2-g/cm\ensuremath{^{\textnormal{2}}} \ensuremath{^{\textnormal{9}}}Be reaction target using the first half of the FRS device that was being operated in separator/spectrometer mode. The Be}\\
\parbox[b][0.3cm]{17.7cm}{reaction target was positioned at the middle focus of the FRS. \ensuremath{^{\textnormal{20}}}Al nuclei, produced via the charge exchange reaction, immediately}\\
\parbox[b][0.3cm]{17.7cm}{decayed to 3p+\ensuremath{^{\textnormal{17}}}Ne products. The second half of the FRS was used to momentum analyze the \ensuremath{^{\textnormal{17}}}Ne heavy decay residues. Their}\\
\parbox[b][0.3cm]{17.7cm}{trajectories were measured at the final focal plane. Decay protons were measured in coincidence (with one another and in}\\
\parbox[b][0.3cm]{17.7cm}{coincidence with the \ensuremath{^{\textnormal{17}}}Ne decay products) using an array of position sensitive Si detectors that were positioned downstream of the}\\
\parbox[b][0.3cm]{17.7cm}{reaction target. Those detectors measured the protons$'$ trajectories.}\\
\parbox[b][0.3cm]{17.7cm}{\addtolength{\parindent}{-0.2in}From \ensuremath{^{\textnormal{17}}}Ne and three protons$'$ trajectories and from 4-fold coincidence events, the angular correlations between each proton and the}\\
\parbox[b][0.3cm]{17.7cm}{\ensuremath{^{\textnormal{17}}}Ne ions, between \ensuremath{^{\textnormal{17}}}Ne-p-p, and between three protons were reconstructed. In addition, the 3p-decay energy was reconstructed.}\\
\parbox[b][0.3cm]{17.7cm}{This analysis reveals that the 3p-unbound \ensuremath{^{\textnormal{20}}}Al nucleus is observed for the first time and indicates that \ensuremath{^{\textnormal{20}}}Al decays via proton}\\
\parbox[b][0.3cm]{17.7cm}{emission to \ensuremath{^{\textnormal{19}}}Mg, which is another proton-unbound nucleus. Two low-lying states at E\ensuremath{_{\textnormal{rel, c.m.}}}(\ensuremath{^{\textnormal{17}}}Ne+3p)\ensuremath{\sim}2.0 MeV and 3.6 MeV}\\
\parbox[b][0.3cm]{17.7cm}{were observed, which are attributed to \ensuremath{^{\textnormal{20}}}Al resonances. The lowest energy state was assumed to be the \ensuremath{^{\textnormal{20}}}Al\ensuremath{_{\textnormal{g.s.}}}. The authors note}\\
\parbox[b][0.3cm]{17.7cm}{that the 3p-decay energy spectrum shows evidence of two higher-energy bumps at E\ensuremath{_{\textnormal{rel, c.m.}}}(\ensuremath{^{\textnormal{17}}}Ne+3p)\ensuremath{\sim}5 MeV and at}\\
\parbox[b][0.3cm]{17.7cm}{E\ensuremath{_{\textnormal{rel, c.m.}}}(\ensuremath{^{\textnormal{17}}}Ne+3p)\ensuremath{\sim}7 MeV, which may correspond to higher excited states of \ensuremath{^{\textnormal{20}}}Al. But those potential states are not analyzed in}\\
\parbox[b][0.3cm]{17.7cm}{this study, and are thus not considered in this evaluation.}\\
\parbox[b][0.3cm]{17.7cm}{\addtolength{\parindent}{-0.2in}Gamow coupled-channel and Gamow shell model calculations were performed to deduce J\ensuremath{^{\ensuremath{\pi}}} values of the observed \ensuremath{^{\textnormal{20}}}Al levels.}\\
\parbox[b][0.3cm]{17.7cm}{The authors discuss isospin symmetry breaking between \ensuremath{^{\textnormal{20}}}Al and \ensuremath{^{\textnormal{20}}}N mirror systems and conclude that the decay energy of \ensuremath{^{\textnormal{20}}}Al}\\
\parbox[b][0.3cm]{17.7cm}{ground state is significantly smaller than the predictions inferred from the isospin symmetry. The reason is most likely due to the}\\
\parbox[b][0.3cm]{17.7cm}{Thomas-Ehrmann shift.}\\
\vspace{12pt}
\underline{$^{20}$Al Levels}\\
\begin{longtable}{cccccccc@{\extracolsep{\fill}}c}
\multicolumn{2}{c}{E(level)$^{}$}&J$^{\pi}$$^{{\hyperlink{AL1LEVEL0}{a}}}$&\multicolumn{2}{c}{\ensuremath{\Gamma}$^{}$}&\multicolumn{2}{c}{E\ensuremath{_{\textnormal{c.m.}}}(\ensuremath{^{\textnormal{17}}}Ne+3p) (MeV)$^{}$}&Comments&\\[-.2cm]
\multicolumn{2}{c}{\hrulefill}&\hrulefill&\multicolumn{2}{c}{\hrulefill}&\multicolumn{2}{c}{\hrulefill}&\hrulefill&
\endfirsthead
\multicolumn{1}{r@{}}{0}&\multicolumn{1}{@{}l}{}&\multicolumn{1}{l}{(1\ensuremath{^{-}})}&\multicolumn{1}{r@{}}{$<$400}&\multicolumn{1}{@{ }l}{keV}&\multicolumn{1}{r@{}}{1}&\multicolumn{1}{@{.}l}{93\ensuremath{^{{\hyperlink{AL1LEVEL1}{b}}}} {\it 16}}&\parbox[t][0.3cm]{8.91939cm}{\raggedright \%p=100 (\href{https://www.nndc.bnl.gov/nsr/nsrlink.jsp?2025Xu03,B}{2025Xu03})\vspace{0.1cm}}&\\
&&&&&&&\parbox[t][0.3cm]{8.91939cm}{\raggedright \ensuremath{\Gamma}: Mainly due to the experimental resolution (see Appendix C\vspace{0.1cm}}&\\
&&&&&&&\parbox[t][0.3cm]{8.91939cm}{\raggedright {\ }{\ }{\ }in \href{https://www.nndc.bnl.gov/nsr/nsrlink.jsp?2025Xu03,B}{2025Xu03}).\vspace{0.1cm}}&\\
&&&&&&&\parbox[t][0.3cm]{8.91939cm}{\raggedright E\ensuremath{_{\textnormal{c.m.}}}(\ensuremath{^{\textnormal{17}}}Ne+3p) (MeV): From S\ensuremath{_{\textnormal{3p}}}(\ensuremath{^{\textnormal{20}}}Al\ensuremath{_{\textnormal{g.s.}}})=1.93 MeV \textit{+12{\textminus}10}\vspace{0.1cm}}&\\
&&&&&&&\parbox[t][0.3cm]{8.91939cm}{\raggedright {\ }{\ }{\ }(\href{https://www.nndc.bnl.gov/nsr/nsrlink.jsp?2025Xu03,B}{2025Xu03}).\vspace{0.1cm}}&\\
&&&&&&&\parbox[t][0.3cm]{8.91939cm}{\raggedright Using S\ensuremath{_{\textnormal{3p}}}(\ensuremath{^{\textnormal{20}}}Al\ensuremath{_{\textnormal{g.s.}}})=1.93 MeV \textit{16} and the mass excesses of\vspace{0.1cm}}&\\
&&&&&&&\parbox[t][0.3cm]{8.91939cm}{\raggedright {\ }{\ }{\ }proton and \ensuremath{^{\textnormal{17}}}Ne from (\href{https://www.nndc.bnl.gov/nsr/nsrlink.jsp?2021Wa16,B}{2021Wa16}), (\href{https://www.nndc.bnl.gov/nsr/nsrlink.jsp?2025Xu03,B}{2025Xu03}) deduced a\vspace{0.1cm}}&\\
&&&&&&&\parbox[t][0.3cm]{8.91939cm}{\raggedright {\ }{\ }{\ }mass excess of 40.30 MeV \textit{12} for \ensuremath{^{\textnormal{20}}}Al.\vspace{0.1cm}}&\\
&&&&&&&\parbox[t][0.3cm]{8.91939cm}{\raggedright S\ensuremath{_{\textnormal{p}}}(\ensuremath{^{\textnormal{20}}}Al\ensuremath{_{\textnormal{g.s.}}})=1.17 MeV \textit{+10{\textminus}8} (\href{https://www.nndc.bnl.gov/nsr/nsrlink.jsp?2025Xu03,B}{2025Xu03}: See Appendix A).\vspace{0.1cm}}&\\
&&&&&&&\parbox[t][0.3cm]{8.91939cm}{\raggedright {\ }{\ }{\ }This value corresponds to a probability of 0.821 from a\vspace{0.1cm}}&\\
&&&&&&&\parbox[t][0.3cm]{8.91939cm}{\raggedright {\ }{\ }{\ }Kolmogorov test.\vspace{0.1cm}}&\\
&&&&&&&\parbox[t][0.3cm]{8.91939cm}{\raggedright Decay mode: p+\ensuremath{^{\textnormal{19}}}Mg\ensuremath{_{\textnormal{g.s.}}}, where \ensuremath{^{\textnormal{19}}}Mg\ensuremath{_{\textnormal{g.s.}}} democratically decays\vspace{0.1cm}}&\\
&&&&&&&\parbox[t][0.3cm]{8.91939cm}{\raggedright {\ }{\ }{\ }to \ensuremath{^{\textnormal{17}}}Ne+2p (\href{https://www.nndc.bnl.gov/nsr/nsrlink.jsp?2025Xu03,B}{2025Xu03}).\vspace{0.1cm}}&\\
&&&&&&&\parbox[t][0.3cm]{8.91939cm}{\raggedright Configuration: \ensuremath{^{\textnormal{19}}}Mg+p, where the proton is predominantly in the\vspace{0.1cm}}&\\
&&&&&&&\parbox[t][0.3cm]{8.91939cm}{\raggedright {\ }{\ }{\ }\textit{s}\ensuremath{_{\textnormal{1/2}}} orbital (\href{https://www.nndc.bnl.gov/nsr/nsrlink.jsp?2025Xu03,B}{2025Xu03}).\vspace{0.1cm}}&\\
\multicolumn{1}{r@{}}{1.67\ensuremath{\times10^{3}}}&\multicolumn{1}{@{ }l}{{\it 27}}&\multicolumn{1}{l}{(2\ensuremath{^{-}})}&&&\multicolumn{1}{r@{}}{3}&\multicolumn{1}{@{.}l}{60 {\it 22}}&\parbox[t][0.3cm]{8.91939cm}{\raggedright \%p=100 (\href{https://www.nndc.bnl.gov/nsr/nsrlink.jsp?2025Xu03,B}{2025Xu03})\vspace{0.1cm}}&\\
&&&&&&&\parbox[t][0.3cm]{8.91939cm}{\raggedright Decays to \ensuremath{^{\textnormal{19}}}Mg*(1.38 MeV, (3/2\ensuremath{^{-}}))+p, where the proton energy\vspace{0.1cm}}&\\
&&&&&&&\parbox[t][0.3cm]{8.91939cm}{\raggedright {\ }{\ }{\ }is deduced to be E\ensuremath{_{\textnormal{c.m.}}}=1.50 MeV \textit{10}. The \ensuremath{^{\textnormal{19}}}Mg*(1.38 MeV)\vspace{0.1cm}}&\\
&&&&&&&\parbox[t][0.3cm]{8.91939cm}{\raggedright {\ }{\ }{\ }level sequentially decays via \ensuremath{^{\textnormal{18}}}Na(g.s., (1\ensuremath{^{-}})) to \ensuremath{^{\textnormal{17}}}Ne\ensuremath{_{\textnormal{g.s.}}}+2p.\vspace{0.1cm}}&\\
\end{longtable}
\parbox[b][0.3cm]{17.7cm}{\makebox[1ex]{\ensuremath{^{\hypertarget{AL1LEVEL0}{a}}}} From Gamow shell model and Gamow coupled-channel model calculations in (\href{https://www.nndc.bnl.gov/nsr/nsrlink.jsp?2025Xu03,B}{2025Xu03}), where the ground and first excited}\\
\parbox[b][0.3cm]{17.7cm}{{\ }{\ }states were firmly and tentatively assigned, respectively. However, the evaluator made the ground state assignment tentative.}\\
\parbox[b][0.3cm]{17.7cm}{\makebox[1ex]{\ensuremath{^{\hypertarget{AL1LEVEL1}{b}}}} Appendix A in (\href{https://www.nndc.bnl.gov/nsr/nsrlink.jsp?2025Xu03,B}{2025Xu03}) presents a minor, alternative 3p-decay channel with the total decay energy E\ensuremath{_{\textnormal{c.m.}}}(\ensuremath{^{\textnormal{17}}}Ne+3p)=1.93}\\
\begin{textblock}{29}(0,27.3)
Continued on next page (footnotes at end of table)
\end{textblock}
\clearpage
\vspace*{-0.5cm}
{\bf \small \underline{\ensuremath{^{\textnormal{9}}}Be(\ensuremath{^{\textnormal{20}}}Mg,\ensuremath{^{\textnormal{20}}}Al)\hspace{0.2in}\href{https://www.nndc.bnl.gov/nsr/nsrlink.jsp?2025Xu03,B}{2025Xu03} (continued)}}\\
\vspace{0.3cm}
\underline{$^{20}$Al Levels (continued)}\\
\vspace{0.3cm}
\parbox[b][0.3cm]{17.7cm}{{\ }{\ }MeV: The decay of the \ensuremath{^{\textnormal{20}}}Al*(1.67 MeV) state may proceed (minor branch) via \ensuremath{^{\textnormal{20}}}Al*\ensuremath{\rightarrow}\ensuremath{^{\textnormal{19}}}Mg*(2.14}\\
\parbox[b][0.3cm]{17.7cm}{{\ }{\ }MeV)+p\ensuremath{\rightarrow}\ensuremath{^{\textnormal{18}}}Na*+2p\ensuremath{\rightarrow}\ensuremath{^{\textnormal{17}}}Ne*(1288)+3p+\ensuremath{\gamma}, where the intermediate state(s) in \ensuremath{^{\textnormal{18}}}Na* are unresolved; this cannot be verified}\\
\parbox[b][0.3cm]{17.7cm}{{\ }{\ }because no \ensuremath{\gamma} rays were measured.}\\
\vspace{0.5cm}
\end{center}
\clearpage
\newpage
\pagestyle{plain}
\section[References]{ }
\vspace{-30pt}
\begin{longtable}{l@{\hskip 0.9cm}l}
\multicolumn{2}{c}{REFERENCES FOR A=20}\\
&\endfirsthead
\multicolumn{2}{c}{REFERENCES FOR A=20(CONTINUED)}\\
&\endhead
\href{https://www.nndc.bnl.gov/nsr/nsrlink.jsp?1987Po01,B}{1987Po01}&\parbox[t]{6in}{\addtolength{\parindent}{-0.25cm}A.Poves, J.Retamosa - Phys.Lett. 184B, 311 (1987).}\\
&\parbox[t]{6in}{\addtolength{\parindent}{-0.25cm} \textit{The Onset of Deformation at the N = 20 Neutron Shell Closure far fromStability.}}\\
\href{https://www.nndc.bnl.gov/nsr/nsrlink.jsp?1997Ki22,B}{1997Ki22}&\parbox[t]{6in}{\addtolength{\parindent}{-0.25cm}H.Kitagawa, N.Tajima, H.Sagawa - Z.Phys. A358, 381 (1997).}\\
&\parbox[t]{6in}{\addtolength{\parindent}{-0.25cm} \textit{Reaction Cross Sections and Radii of A = 17 and A = 20 Isobars.}}\\
\href{https://www.nndc.bnl.gov/nsr/nsrlink.jsp?1999Kn04,B}{1999Kn04}&\parbox[t]{6in}{\addtolength{\parindent}{-0.25cm}O.M.Knyazkov, I.N.Kukhtina, S.A.Fayans - Fiz.Elem.Chastits At.Yadra 30, 870 (1999); Phys.Part.Nucl. 30, 369 (1999).}\\
&\parbox[t]{6in}{\addtolength{\parindent}{-0.25cm} \textit{Interaction Cross Sections and Structure of Light Exotic Nuclei.}}\\
\href{https://www.nndc.bnl.gov/nsr/nsrlink.jsp?2007Mu15,B}{2007Mu15}&\parbox[t]{6in}{\addtolength{\parindent}{-0.25cm}I.Mukha, K.Summerer, L.Acosta, M.A.G.Alvarez et al. - Phys.Rev.Lett. 99, 182501 (2007).}\\
&\parbox[t]{6in}{\addtolength{\parindent}{-0.25cm} \textit{Observation of Two-Proton Radioactivity of \ensuremath{^{\textnormal{19}}}Mg by Tracking the Decay Products.}}\\
\href{https://www.nndc.bnl.gov/nsr/nsrlink.jsp?2011Pr03,B}{2011Pr03}&\parbox[t]{6in}{\addtolength{\parindent}{-0.25cm}B.Pritychenko, E.Betak, M.A.Kellett, B.Singh, J.Totans - Nucl.Instrum.Methods Phys.Res. A640, 213 (2011).}\\
&\parbox[t]{6in}{\addtolength{\parindent}{-0.25cm} \textit{The Nuclear Science References (NSR) database and Web Retrieval System.}}\\
\href{https://www.nndc.bnl.gov/nsr/nsrlink.jsp?2013Ti01,B}{2013Ti01}&\parbox[t]{6in}{\addtolength{\parindent}{-0.25cm}J.Tian, N.Wang, C.Li, J.Li - Phys.Rev. C 87, 014313 (2013).}\\
&\parbox[t]{6in}{\addtolength{\parindent}{-0.25cm} \textit{Improved Kelson-Garvey mass relations for proton-rich nuclei.}}\\
\href{https://www.nndc.bnl.gov/nsr/nsrlink.jsp?2018Fo04,B}{2018Fo04}&\parbox[t]{6in}{\addtolength{\parindent}{-0.25cm}H.T.Fortune - Phys.Rev. C 97, 034301 (2018).}\\
&\parbox[t]{6in}{\addtolength{\parindent}{-0.25cm} \textit{Mirror energy differences of 2s\ensuremath{_{\textnormal{1/2}}}, 1d\ensuremath{_{\textnormal{5/2}}} and 1f\ensuremath{_{\textnormal{7/2}}} states.}}\\
\href{https://www.nndc.bnl.gov/nsr/nsrlink.jsp?2021Wa16,B}{2021Wa16}&\parbox[t]{6in}{\addtolength{\parindent}{-0.25cm}M.Wang, W.J.Huang, F.G.Kondev, G.Audi, S.Naimi - Chin.Phys.C 45, 030003 (2021).}\\
&\parbox[t]{6in}{\addtolength{\parindent}{-0.25cm} \textit{The AME 2020 atomic mass evaluation (II). Tables, graphs and references.}}\\
\href{https://www.nndc.bnl.gov/nsr/nsrlink.jsp?2022Zo01,B}{2022Zo01}&\parbox[t]{6in}{\addtolength{\parindent}{-0.25cm}Y.Y.Zong, C.Ma, M.Q.Lin, Y.M.Zhao - Phys.Rev. C 105, 034321 (2022).}\\
&\parbox[t]{6in}{\addtolength{\parindent}{-0.25cm} \textit{Mass relations of mirror nuclei for both bound and unbound systems.}}\\
\href{https://www.nndc.bnl.gov/nsr/nsrlink.jsp?2025Xu03,B}{2025Xu03}&\parbox[t]{6in}{\addtolength{\parindent}{-0.25cm}X.-D.Xu, I.Mukha, J.G.Li, S.M.Wang et al. - Phys.Rev.Lett. 135, 022502 (2025).}\\
&\parbox[t]{6in}{\addtolength{\parindent}{-0.25cm} \textit{Isospin Symmetry Breaking Disclosed in the Decay of Three-Proton Emitter \ensuremath{^{\textnormal{20}}}Al.}}\\
\end{longtable}
\end{document}